\documentclass[11pt,twoside]{article}
\usepackage{asp2010}

\resetcounters

\begin{document}

\title{Zeeman Broadening in Optical Stokes I of Solar-like Stars}
\author{Richard I. Anderson$^{1,2}$, Ansgar Reiners$^2$, and Sami K. Solanki$^{3,4}$
\affil{$^1$Observatoire de Gen\`eve, Universit\'e de Gen\`eve, 51 Ch. des Maillettes, CH-1290 Sauverny, Switzerland}
\affil{$^2$Institut f\"ur Astrophysik, Georg-August-Universit\"at
    G\"ottingen, Friedrich-Hund-Platz 1, D-37077 G\"ottingen,
    Germany}
\affil{$^3$Max-Planck-Institut f\"ur Sonnensystemforschung,
    Max-Planck-Stra\ss e 2, D-37191 Katlenburg-Lindau, Germany}
\affil{$^4$School of Space Research, Kyung Hee University, Yongin, Gyeonggi 446-701, Korea}}

\begin{abstract}
We present our analysis aimed at inferring average magnetic fields in
slowly-rotating solar-like stars.
Using the spectral line inversion code \texttt{SPINOR}, we perform
high-accuracy line profile fitting and investigate whether Zeeman
broadening can be reliably detected in optical data of unprecedented
quality. 
We argue that our usage of both high- and low-$g_{\rm{eff}}$ lines
does provide a certain sensitivity to magnetic fields that
may, indeed, be detected. 
However, the measurement is subject to a model dependence and prone to
ambiguities, e.g. due to spectral blends.
Hence, while a field may be successfully recovered, the quantification
of this field is subject to large uncertainties, even for the
highest-quality optical data. 
\end{abstract}

\section{Introduction}
Direct detections of magnetic fields in slowly-rotating ($v\sin i <
10\,\rm{km\,s}^{-1}$) solar-like stars (SRSLS) are somewhat
  rare. Since the first application of the Zeeman broadening 
technique to cool stars by \cite{1980ApJ...239..961R}, much effort was
made to infer their field strengths, e.g. in order to constrain stellar dynamo 
theory. The more recent advances, however, come from different
directions, such as direct detections for very late-type stars,
e.g. \cite{2007ApJ...656.1121R}, and detections of 
magnetic field geometry reversals for faster rotators
\citep{2009MNRAS.398.1383F}. The rotation-activity relationship and
the absence of accurate and recent measurements of SRSLS
magnetic fields thus suggest that the corresponding field strengths
may be too low to enable clear and robust detections.  

We aim to shed some light on the question of Zeeman broadening
detectability in optical Stokes\,I of SRSLS, and quantify the
fields found. To this end, we apply the
state-of-the-art line inversion code \texttt{SPINOR}
\citep[][]{2005A&A...444..549F} to data of unprecedented quality 
and search for signs of Zeeman broadening, see
\cite{2010A&A...522A..81A}[from hereon PaperI].

\section{Observations and Analysis}\label{Sec:Analysis}
The spectra used in this work were taken with the Coud\'e-\'Echelle
Spectrometer, formerly fiber-fed from the ESO 3.6m at La Silla, Chile.   
The resolving power $R = 220\,000$ and S/N between $400$
and $700$ render these the best data considered in this kind of
analysis as of yet. This high resolution, however, comes at the
expense of a very short wavelength coverage of only $40$\,\AA,
severely limiting the number of spectral lines suitable for
analysis. The main criterion for suitability is the absence of 
blends (or at least that the blends present can be well enough
reproduced), a utopia as becomes clear with increasing data quality. 

Zeeman splitting is a second-order line broadening effect (the
splitting of the circularly polarized $\sigma$ components is not
resolved) in the optical range and is easily outweighed by thermal,
rotational, or turbulent broadening.  However, contrary to the latter,
Zeeman splitting depends on $\lambda^2$.  
Additionally, there exists a linear dependence on the Land\'e factor
of the transition, usually expressed in terms of the effective Land\'e
factor, $g_{\rm{eff}}$.  
These dependences have two important consequences:  
\begin{enumerate}
\item Zeeman splitting can become stronger than e.g. rotational line
  broadening in slowly rotating stars in the infrared \label{item:one}
\item the extent of magnetic broadening varies between lines of
  different $g_{\rm{eff}}$ and thereby provides a way of
  distinguishing between Zeeman broadening and other broadening
  agents. \label{item:two}
\end{enumerate}
As we use optical spectra, we do not benefit from
point\,\ref{item:one}. However, due to the very high precision of CES,
we can attempt to exploit point\,\ref{item:two} by using both high- and
low-$g_{\rm{eff}}$ lines (cf. Tab.\,\ref{tab:lines}) in simultaneous
inversions. 
Thereby, we have a strong magnetic field discriminant at our disposal
whose detectability limit we can explore.
\begin{table}[!b]
\caption{Lines used in inversions. Oscillator
  strengths, abundances adopted, and blends included in calculation of
  line profiles presented in PaperI. $g_{\rm{eff}}$ is the effective
  Land\'e factor, $\chi_e$ the lower excitation potential.}
\label{tab:lines}
\smallskip
\begin{center}
{\small
\begin{tabular}{cccccccccc}
\tableline
\noalign{\smallskip}
\multicolumn{4}{c}{Data set A: $5770\,\AA$} & & & \multicolumn{4}{c}{Data set B: $6137\,\AA$} \\
$\lambda\,[\AA]$ & Ion & $g_{\rm{eff}}$ & $\chi_e$ [eV] & & &
$\lambda\,[\AA]$ &Ion & $g_{\rm{eff}}$ & $\chi_e$ [eV] \\ 
\noalign{\smallskip}
\tableline
\noalign{\smallskip}
5778.46 & Fe\,I & 1.21 & 2.59 & & & 6151.62 & Fe\,I & 1.83 & 2.18  \\ 
5783.06 & Cr\,I & 2.00 & 3.32 & & & 6165.36 & Fe\,I & 1.00 & 4.14  \\
5806.73 & Fe\,I & 0.92 & 4.61 & & & 6173.34 & Fe\,I & 2.50 & 2.22  \\
\noalign{\smallskip}
\tableline
\end{tabular}
}
\end{center}
\end{table}

We investigate two kinds of models, one-component (OC) and
two-component (TC) models. The two differ in surface field coverage;
OC models assume a surface covered entirely by an average magnetic
field, while TC models distinguish between the magnetic and
non-magnetic surface fractions. The filling factor $f$ represents the
magnetic surface fraction. In TC models, we investigate the impact of
different assumptions on the temperatures of both components. 

A grid of fixed values for $B$ and $f$ (in OC models, $f$ is effectively $1$)
is provided. Best-fit values for the merit function and the free
parameters (temperature, $v\sin i$, and turbulent
velocities) are computed using a non-linear $\chi^2$ 
minimization algorithm \citep{Frutithesis}. To claim detection of
Zeeman broadening, we require that the $3\,\sigma$ confidence level
\citep[$3\,\sigma$\,CL, defined by $\chi^{2}_{\rm{min}}+9$,
  see][]{1992nrca.book.....P} be inconsistent with $Bf = 0\,G$.  More
detail on the analysis and the inversion code \texttt{SPINOR}, 
can be found in PaperI and references therein. 

\section{Results}
\begin{figure}[!ht]
\plotone{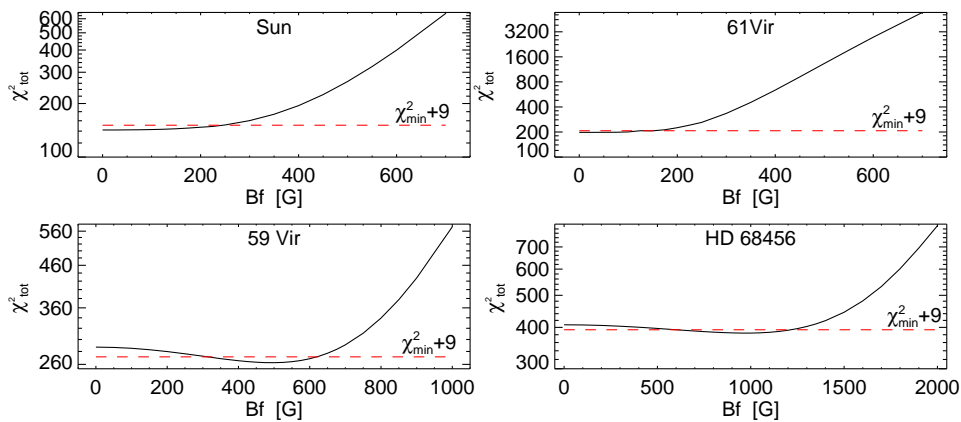}
\caption{Some OC results shown in PaperI, Fig.\,3: logarithmically-drawn
  best-fit $\chi^2$ against fixed $Bf$ for the Sun, 61\,Vir,
  59\,Vir, and HD\,68456. $3\,\sigma$ CLs indicated by dashed lines.}
\label{fig:chimaps}
\end{figure}

Figure\,\ref{fig:chimaps} shows the behavior of $\chi^2$ as a function
of fixed $Bf$ values in OC 
inversions of (top left to bottom right) an integrated solar
spectrum (sunlight reflected by Ganymede, close to maximum activity in
October 2000), the inactive G6 dwarf 61\,Vir, the active G0 dwarf 
59\,Vir, and the active F6 dwarf HD\,68456; spectral types taken from
\texttt{SIMBAD}, active denotes high X-ray luminosity as found in the
NEXXUS2 database  \citep[][\url{www.hs.uni-hamburg.de/DE/For/Gal/Xgroup/nexxus/nexxus.html}]{2004A&A...417..651S} .

In Tab.\,\ref{tab:stars}, we list the results for the stars from
Fig.\,\ref{fig:chimaps} together with their logarithmic X-ray
luminosities, projected rotational velocities, and the results
obtained for our OC inversions of data set A for the Sun and data set
B for 61\,Vir, 59\,Vir, and HD\,68456. 

\begin{table}[!t]
\caption{Stars investigated with their activity-relevant
  parameters. $\log{L_X}$ from \cite{2003ApJ...593..534J} (Sun) and
  NEXXUS2 (stars), $v\sin i$ from inversions (consistent
  with literature), the $3\,\sigma$ CLs on $Bf$ are from OC inversions.
}
\label{tab:stars}
\smallskip
\begin{center}
{\small
\begin{tabular}{rrcccc}
\tableline
\noalign{\smallskip}
HD Number & alt. ID & Sp.Type & $\log{L_X}$ & $v\sin i$ &
$Bf\,3\,\sigma$-range\\ 
   & & &  [ergs s$^{-1}$] & [km s$^{-1}$]& [$G$] \\
\noalign{\smallskip}
\tableline
\noalign{\smallskip}
 & Sun (active) & G2V & $27.1 - 27.75$ & $1.3$ & $0 - 200$ \\\noalign{\smallskip}
68456 & GJ 297.1 & F6V & $29.05$ & $9.5$ & $600 - 1200$ \\\noalign{\smallskip}
115383 & 59\,Vir & G0V & $29.41$ & $6.7$ & $300 - 600$ \\\noalign{\smallskip}
115617 & 61\,Vir & G6V & $26.65$ & $0.5$ & $0 - 150$ \\
\noalign{\smallskip}
\tableline
\end{tabular}
}
\end{center}
\end{table}

OC inversions clearly exclude significant magnetic flux ($> 200\,G$)
on the (active) Sun and $61$\,Vir. Under the OC assumption, we find
significant magnetic flux for 59\,Vir and HD\,68456. 
For 59\,Vir, TC inversions yielded results consistent with both the OC
result and $Bf = 0$\,G at the $3\,\sigma$\,CL, since $Bf$ was
$0-300$\,G (cool magnetic regions case). Comparing the three 
cases of equal temperature, warm, and cool magnetic regions, 
we find that the influence on $\chi^2$ of a second temperature
component is considerably larger than that of even a rather strong
magnetic field. Hence, the numerical value of the OC detection is
challenged. However, the TC results for 59\,Vir are inconsistent 
with $Bf = 0\,G$ at the $2\,\sigma$\,CL.   

\section{Conclusion}
Our analysis to infer average magnetic fields in optical spectra takes
us close to the detectability limit of Zeeman broadening for SRSLS. 
Overall, we find that even the highest quality data
do not yield clear and robust detections. The analysis is subject to a
model dependence related to the inital choice of temperature
configuration for the stellar surface, and may further be influenced by
the approximative treatment of convection by analytical macroturbulence
profiles (see PaperI). Also, blends (and their reproducability) gain in 
importance at the present level of accuracy and therefore constitute a
limiting factor. However, there does remain a sensitivity to the
Zeeman broadening signature, cf. the case of 59\,Vir, thanks to
the simultaneous use of both high- and low-$g_{\rm{eff}}$
lines. It is therefore the numerical amount of magnetic flux detected
that is questioned by the OC-TC-related model dependence, not the
presence of a magnetic field on the star.  

\acknowledgements RIA and AR acknowledge research funding from the DFG
  under an Emmy Noether Fellowship (RE 1664/4-1). RIA further acknowledges
  funding by the Fonds National Suisse de la Recherche Scientifique
  (FNRS).  The work of SKS has been partially supported by the
  WCU grant No. R31-10016 funded by the Korean Ministry of Education, Science
  and Technology.
  This research has made use of NASA's Astrophysics Data System
  Bibliographic Services and the SIMBAD database,
  operated at CDS, Strasbourg, France. 

\bibliography{anderson_r}

\end{document}